\documentclass[%
 reprint,
 superscriptaddress,
 amsmath,amssymb,
 aps,
showkeys,
]{revtex4-2}

\usepackage{graphicx}
\usepackage{dcolumn}
\usepackage{bm}
\usepackage{comment}
\usepackage{xcolor}

\usepackage[utf8]{inputenc}
\usepackage{amsmath} 
\usepackage{xspace}	

\newcommand\nm{~{\rm nm}}
\newcommand\um{~{\rm \mu m}}

\newcommand\mm{~{\rm mm}}

\newcommand{\cmqW}{~{\rm cm^2/W}}
\newcommand{\figname}{Fig.~}

\newcommand{\fignametot}{Figure~}
\newcommand{\appname}{Appendix~}
\newcommand{\eqname}{Eq.~}
\newcommand{\forward}{\textit{forward}\xspace}

\newcommand{\reverse}{\textit{reverse}\xspace}

\newcommand{\FP}{FP\xspace}
\newcommand{\FPO}{FPO\xspace}
\newcommand{\FPOs}{FPOs\xspace}
\newcommand{\waveguide}{BW\xspace} 

\begin{document}


\title{Influence of the bus waveguide on the linear and nonlinear response of a taiji microresonator}

\author{Riccardo Franchi}
    \email{corresponding author: riccardo.franchi@unitn.it}
    \affiliation{Nanoscience Laboratory, Department of Physics, University of Trento, Via Sommarive 14, Povo - Trento, 38123, Italy}
\author{Stefano Biasi}%
    \affiliation{Nanoscience Laboratory, Department of Physics, University of Trento, Via Sommarive 14, Povo - Trento, 38123, Italy}
\author{Alberto Muñoz de las Heras}
    \affiliation{INO-CNR BEC Center and Department of Physics, University of Trento, 38123 Trento, Italy}
\author{Mher Ghulinyan}
    \affiliation{Sensors and Devices, Fondazione Bruno Kessler, 38123 Trento, Italy}
\author{Iacopo Carusotto}
    \affiliation{INO-CNR BEC Center and Department of Physics, University of Trento, 38123 Trento, Italy}
\author{Lorenzo Pavesi}
    \affiliation{Nanoscience Laboratory, Department of Physics, University of Trento, Via Sommarive 14, Povo - Trento, 38123, Italy}

\date{\today}

\begin{abstract}
We study the linear and nonlinear response of a unidirectional reflector where a nonlinear breaking of the Lorentz reciprocity is observed. The device under test consists  of a racetrack microresonator, with an embedded S-shaped waveguide, coupled to an external bus waveguide (\waveguide). This geometry of the microresonator, known as ``taiji'' microresonator (TJMR), allows to selectively couple counter-propagating modes depending on the propagation direction of the incident light and, at the nonlinear level, leads to an effective breaking of Lorentz reciprocity. Here, we show that a full description of the device needs to consider also the role of the \waveguide, which introduces (i) Fabry-Perot oscillations (\FPOs) due to reflections at its facets, and (ii) asymmetric losses, which depend on the actual position of the TJMR. At sufficiently low powers the asymmetric loss does not affect the unidirectional behavior, but the \FP interference fringes can cancel the effect of the S-shaped waveguide. However, at high input power, both the asymmetric loss and the \FPOs contribute to the redistribution of the energy between the clockwise and counterclockwise modes within the TJMR. This strongly modifies the nonlinear response, giving rise to counter-intuitive features where, due to the \FP effect and the asymmetric losses, the \waveguide properties can determine the violation of the Lorentz reciprocity and, in particular, the difference between the transmittance in the two directions of excitation. The experimental results are explained by using an analytical model based on the transfer matrix approach, a numerical finite-element model and exploiting intuitive interference diagrams.
\end{abstract}



\maketitle

\section{Introduction}

In the last decade, several efforts have been spent to implement optical circuits which show different behavior depending on the propagation direction of the incident light \cite{Fan_2012, Li_2020, Aleahmad_2017, Peng_2014, Chang_2014, DelBino_2017, DelBino_2018, Alberto2021, Dotsch_2005, Wang_2005, Wang_2009,Bi_2011, Shoji_2012, Ozawa2019, Yan_2020, Hohimer_1993, Kharitonov_2015,Ren_2018, Bandres_2018, Smirnova2020}.
The realization of an integrated system capable of working as an optical isolator in the linear regime is prohibited by the Lorentz reciprocity theorem \cite{Potton_2004, Chew2008}.
This ensures that transmission through any linear and non-magnetic media does not depend on the direction of propagation.
However, by properly engineering the optical system, it is possible to induce a non-Hermitian behavior and obtain direction-dependent properties \cite{Wiersig2020, Ganainy2019, Calabrese2020}.
A widely exploited non-Hermitian system is a racetrack microresonator with an embedded S-shaped waveguide (taiji microresonator, TJMR).
The TJMR with a gain medium has been studied to achieve unidirectional behaviour in semiconductor ring laser devices \cite{Hohimer_1993,Kharitonov_2015,Ren_2018} and, recently, in topological lasers \cite{Bandres_2018,Maczewsky_2020}.
In \cite{Calabrese2020}, we studied the unidirectional reflector behaviour of TJMR. When a TJMR is coupled to a bus waveguide (\waveguide), the transmission in both excitation directions is the same while the reflection can assume completely different values. Moreover, such a non-Hermitian design can be combined with the nonlinear material response to break the Lorentz reciprocity theorem, as was demonstrated in \cite{Alberto2021}. There, the breaking of the reciprocity is observed in both a direction-dependent nonlinear shift of the TJMR resonances as well as in a direction-dependent optical bistability loop. These results are strictly related to the role of the S-shaped waveguide which allows to selectively couple the counter-propagating modes in a direction-dependent way. Therefore, the energy stored within the TJMR shows different values for the different excitation directions. 

While the experiments in~\cite{Alberto2021} were restricted to the simplest configurations and provided a pioneering understanding on the basic effect, here we proceed in our analysis by investigating in full detail the role of the \waveguide in this physics. In fact, the reflections at the \waveguide facets\cite{Pruessner2007, Barrios2004} and the \waveguide propagation losses cause a redistribution of the internal energy in the TJMR which depends on the actual position of the microresonator along the waveguide. 
In particular, we report a joint experimental and theoretical study of the interference of the \waveguide optical mode with the clockwise (CW) and counterclockwise (CCW) TJMR modes. We discuss the response of the system in the linear and nonlinear regimes where the microresonator-position-dependent asymmetric propagation losses and the Fabry-Perot oscillations (\FPOs) redistribute the internal energy between both modes yielding a direction-dependent response.

The structure of the paper is the following. In section \ref{Sec:Exp} we report the experimental evidences of different transmission and reflection behaviors in the linear and nonlinear regimes. In section \ref{Sec:Theory} we discuss the numerical simulations which reproduce the experimental observation. In section \ref{Sec:Conc} we draw the conclusions.

\section{Experiments}
\label{Sec:Exp}
\subsection{The device and the experimental setup}
\begin{figure}[t!]
    \centering
    \includegraphics[width=0.45\textwidth]{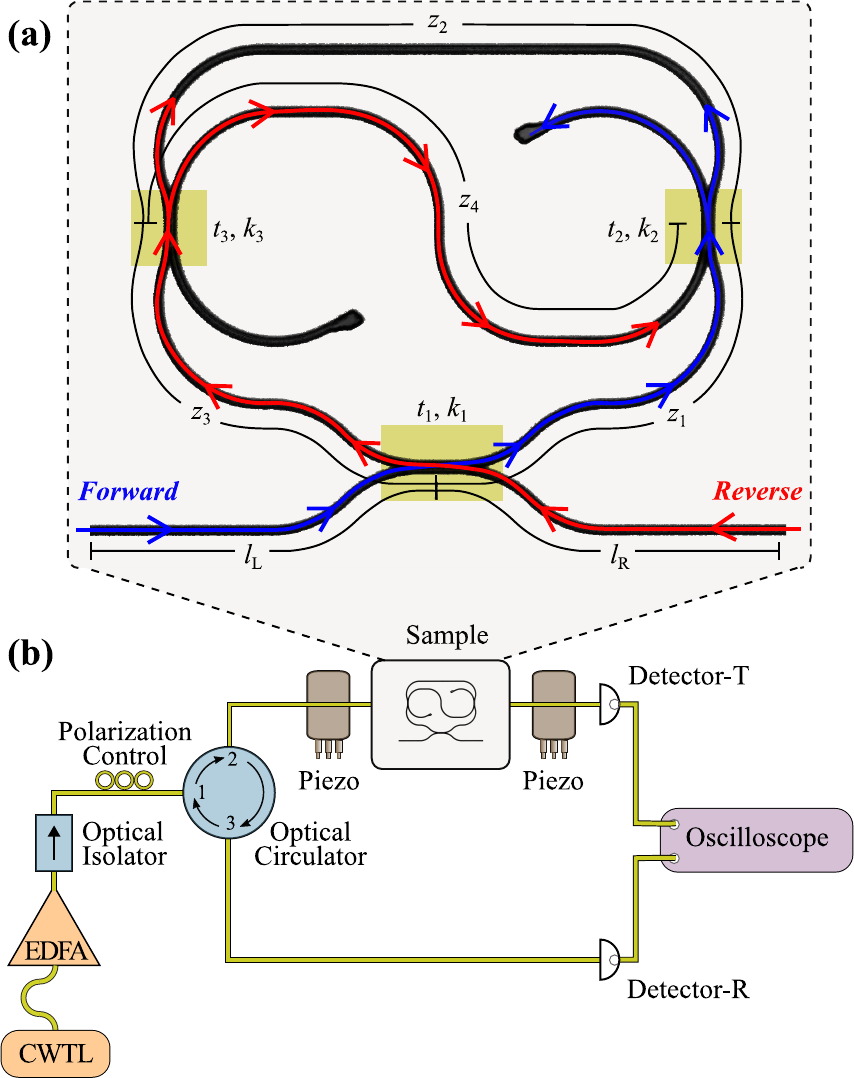}
    \caption{(a) Design of the TJMR coupled to the \waveguide. The yellow rectangles show the three directional couplers. The blue and red arrows highlight the light path in the \forward and \reverse configurations. (b) Sketch of the experimental setup. CWTL: Continuous Wave Tunable Laser, EDFA: Erbium Doped Fiber Amplifier.}
    \label{fig:Setup}
\end{figure}

The \waveguide/TJMR coupled system (the device in the following) is built on single mode channel waveguides made of a silicon oxynitride (SiON) film on a 6 inch Silicon wafer, see \cite{Calabrese2020} for more details. The TJMR consists of a racetrack resonator with a S-shaped waveguide across, as shown in \figname\ref{fig:Setup} (a).
The tips of the S-shaped waveguide have a dark cavity shape to trap the propagating mode and, consequently, to avoid back-reflections \cite{Castellan_2016}.
The coupling between the waveguides is ensured by three directional couplers: one for the \waveguide ($t_1\, ,k_1$) and two for the S-shaped branch ($t_2\, ,k_2$ and $t_3\, ,k_3$).
The perimeter of the racetrack is defined as $p=z_1+z_2+z_3=810.24\um$ (see \figname\ref{fig:Setup}), while the length of the S-shaped waveguide is $z_4=391.12\um$. 
The \waveguide has two polished end facets where light is input or output by butt coupling tapered fibers. Its length is given by  $l_{\rm L}+l_{\rm R}$.  $l_{\rm L}$ and $l_{\rm R}$ define the relative position of the TJMR along the \waveguide.
We measured two samples with equal TJMR parameters and $l_{\rm L} \simeq 0.431\mm$ while different $l_{\rm R} \simeq 5.52\mm$ and $l_{\rm R} \simeq 1.062\mm$. More details on the device are reported in \cite{Calabrese2020}.

The experimental setup shown in \figname\ref{fig:Setup} allows measuring the transmission and reflection spectra of the device. 
A continuous wave tunable laser (Yenista OPTICS, TUNICS-T100S) operating in the IR range ($1490$ - $1640\nm$) is fiber-coupled to an erbium doped fiber amplifier (IPG photonics). In order to prevent laser damage its emission passes through an isolator and the resulting signal is adjusted in polarization by means of a polarization control stage. After it, the light is coupled to a fiber-circulator, which sends the light into a lensed tapered fiber. The light is then butt-coupled to the device using a xyz piezo-positioner for a correct alignment. At the device output, another lensed tapered fiber collects the transmission response and sends the light into an InGaAs detector-T (Thorlabs, PDA20CS(-EC)).
At the same time, the light, which is back-reflected by the device input facet, is filtered out by the circulator and it is acquired by another InGaAs detector-R (Thorlabs, PDA20CS2). The detector-T and detector-R signals are then measured simultaneously with an oscilloscope (PicoScope 4000 Series). We note that at high input powers, only the transmission spectra are measured because of the damage threshold of the optical circulator. 

Turning the device on the sample holder, we input the light in either \forward or in  \reverse configurations. In the \forward configuration, light is CCW-coupled to the TJMR (see blue arrows in \figname\ref{fig:Setup} (a)). Therefore, neglecting the \FPOs due to reflections at the \waveguide facets, the light circulates into the outer path and the S-shaped waveguide is just a source of losses. In this case only a finite transmittance is recorded. In the \reverse configuration, light is CW-coupled to the TJMR and part of it is coupled into the CCW direction by means of the S-shaped waveguide (see red arrows in \figname\ref{fig:Setup} (a)). Therefore, in this case, we do measure both finite transmission and reflection signals from the device. In the linear regime this leads to the unidirectional reflector behavior described in Ref.~\cite{Calabrese2020}.

\subsection{Experimental results in the linear regime}
\label{subsec:ExpLinearRegim}

\begin{figure}[t!]
	\centering
	\includegraphics[width=0.45\textwidth]{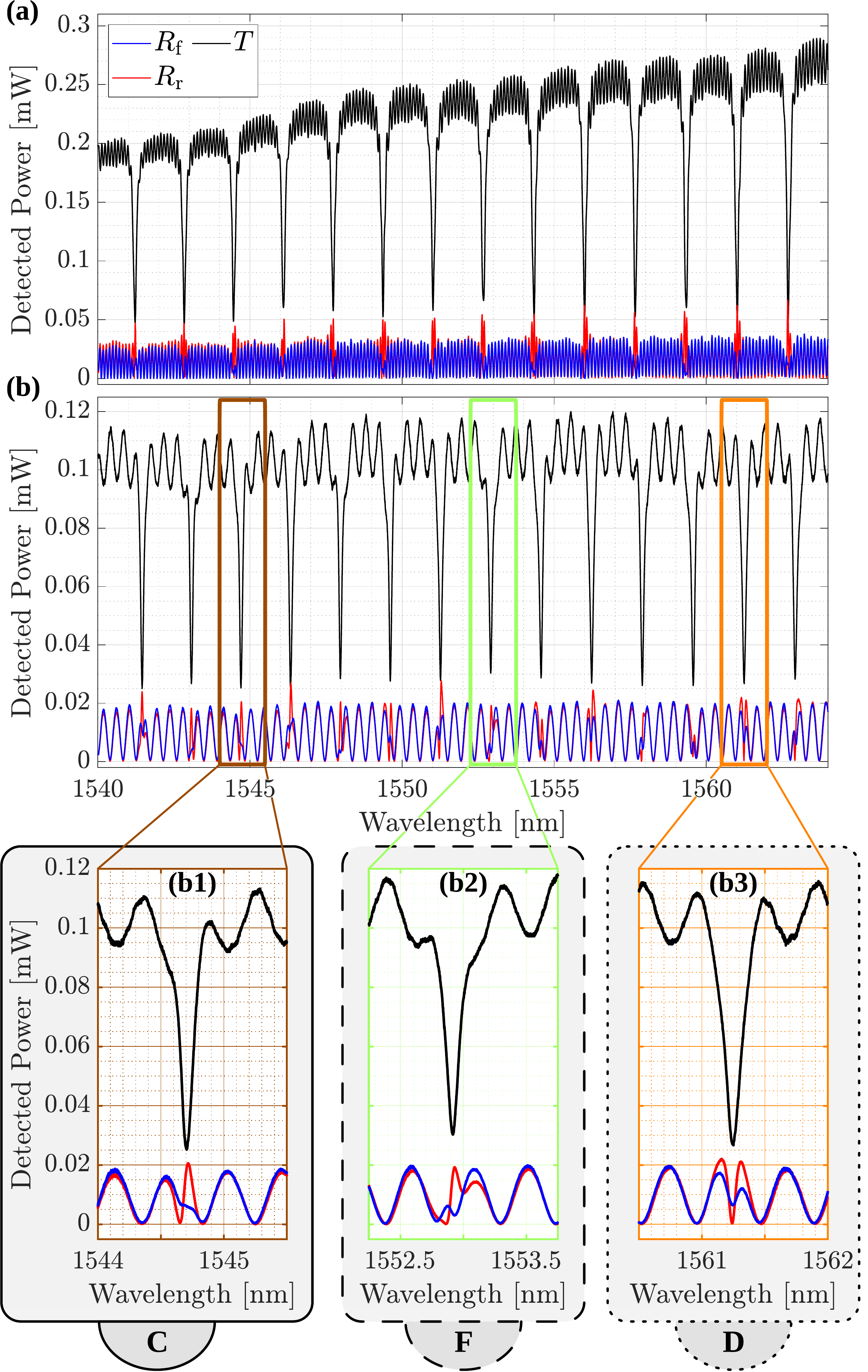}
	\caption{Transmission (black lines) and reflection spectra (blue lines for the \forward configuration and red lines for \reverse configuration) for a device with $l_{\rm R}\simeq 5.52\mm$ (a) and with $l_{\rm R}\simeq 1.062\mm$ (b). The zooms (b1), (b2) and (b3) highlight three different cases, where interference gives rise to different lineshapes: \textit{constructive}-like (C), \textit{Fano}-like (F) and \textit{destructive}-like (D) shapes, respectively.}
	\label{fig:LinearResponse}
\end{figure}

The transmission and reflection spectra of two devices with a different \waveguide length are shown in \fignametot\ref{fig:LinearResponse} for both \forward and \reverse configurations.
Panel (a) refers to $l_{\rm R}\simeq 5.52\mm$  while panel (b) to $l_{\rm R}\simeq 1.062\mm$.
In agreement with the Lorentz reciprocity theorem, the transmission spectra for the \forward and \reverse configurations are the same. They display a set of resonance dips at the TJMR resonances within short \FPOs due to the reflection at the input and output facets of the \waveguide.
Each resonance dip exhibits a typical Lorentzian shape and never shows a doublet as in the case of backscattering \cite{Biasi2019, Li2016}. This means that, in our case, the surface wall roughness is not a dominant source of intrinsic losses. Consequently, it does not contribute to the non-Hermitian dynamics induced by the presence of embedded S-shaped \cite{Calabrese2020}. It is worth noting that out of resonance the two reflections overlaps perfectly. Therefore, the two bus waveguide facets contribute in the same fashion to the reflected component of the optical field.
As expected, by decreasing the \waveguide length, the number of resonances remains constant, while the \FP period increases by about four times (see \figname\ref{fig:LinearResponse} (b)).
In the \forward orientation, this variation does not modify the reflection response of the device which shows the usual \FP fringes (see blue curves of \figname\ref{fig:LinearResponse}). On the other hand, the reflection in the \reverse configuration drastically changes. Specifically, in panel (a) the reflected intensity always shows clear resonance peaks, while in panel (b) it strongly varies as a function of the incident wavelength. This is due to the fact that the short interference fringes of the long device do not affect the reflected optical mode from the TJMR while the long \FP interference fringes in the short device cause significant interference between the taiji reflected mode and the \waveguide modes. This interference may destroy the effect of the S-shaped waveguide in the device reflection.
Specifically, as shown in the zoom of \figname\ref{fig:LinearResponse} (b), we observe three main cases: \textit{constructive}-like (b1), \textit{Fano}-like (b2) and \textit{destructive}-like (b3) reflection lineshape.
In the first case (denoted with the letter C), constructive interference generates a resonant peak and, therefore, the device behaves as the typical TJMR \cite{Calabrese2020}. In the second case (denoted with the letter F), the interference gives rise to a sharp peak with the same height of the \FPO. Interestingly, in the third case (denoted with the letter D), destructive interference rules out the resonant reflection peak. In this case, the efficiency of the taiji as an unidirectional reflection device is much reduced.

\subsection{Experimental results in the nonlinear regime}
\label{subsec:ExpNonLinearRegim}
The three interference cases described in \ref{subsec:ExpLinearRegim} affect also the nonlinear response of the device. As demonstrated in \cite{Alberto2021}, the TJMR exhibits a higher internal power in the \reverse than in the \forward configuration. In fact, in the \forward configuration, the light is partially lost at the end of the S-shaped waveguide. On the other hand, in the \reverse one, the S-shaped branch couples light from the CW to the CCW mode increasing the stored energy. As a result, the transmission response of the device to strong fields shows a non-reciprocal behavior. Since the reflected intensity is strictly connected to the energy stored inside the taiji, the \FP and the propagation losses of the \waveguide strongly affect the nonlinearity-induced non-reciprocal response. 

First, we studied the role of the \FP. We measured the transmitted spectra for different input powers ($P$). 
\begin{figure}[t!]
    \centering
    \includegraphics[width=0.45\textwidth]{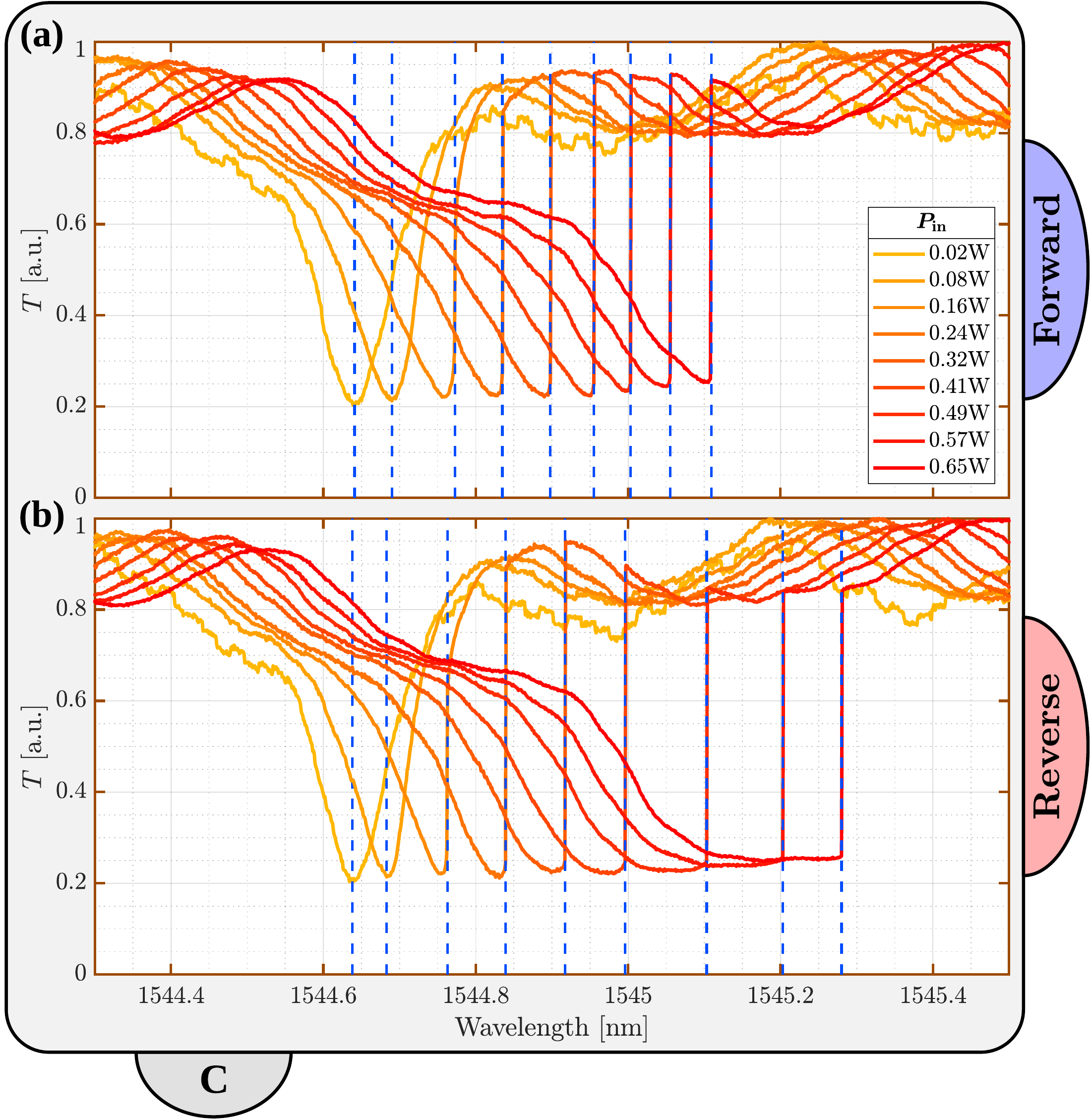}
    \caption{Normalized transmission spectra of upwards wavelength ramps for a device, which shows a \textit{constructive}-like case in the linear regime, for different input powers. The wavelength is scanned from low to high values. Panel (a) and (b) show, respectively, the \forward and \reverse orientations. The blue dashed lines highlight the resonant wavelengths for the different powers. At low input power, this is taken as the minimum of the Lorentzian dips, while at high input power, this is taken as the wavelength at which the transmission jump occurs.}
    \label{fig:NonLinearCostructive}
\end{figure}
\fignametot\ref{fig:NonLinearCostructive} shows the transmission in \forward (Fig. \ref{fig:NonLinearCostructive} (a)) and in \reverse (Fig. \ref{fig:NonLinearCostructive} (b)) configurations for a resonance showing \textit{constructive}-like feature in reflection in the linear regime.
At low $P$, the device exhibits the same resonance Lorentzian dips for both orientations. Increasing $P$, the resonance is pushed towards longer wavelengths due to the build-up of the internal energy in the TJMR and the thermo-optic nonlinearity, see \appname\ref{app:model}. Also, the lineshape changes and takes the typical triangular shape of a microresonator under strong pumping \cite{Ilchenko_1992, Carmon2004, Schmidt2008, RamiroManzano_2013}. To quantify these behavior we trace the resonance wavelength ($\lambda^{\rm res}$) as a function of $P$. As the \FPOs modify the wavelength at which transmittance reaches its minimum value, $\lambda^{\rm res}$ is measured as the wavelength position of the transmission dip at low $P$, and as the threshold wavelength at which the transmittance switches to a value close to one after optical bistability \cite{Boyd, Butcher, Gibbs_1985, Ilchenko_1992, Carmon2004, Almeida2004, RamiroManzano_2013, Trenti_2018, Ghalanos_2020} for the higher values of $P$. Note that a large stored energy in the TJMR gives rise to a larger $\lambda^{\rm res}$ shift, as show in \cite{Alberto2021}.

\begin{figure}[t!]
    \centering
    \includegraphics[width=0.45\textwidth]{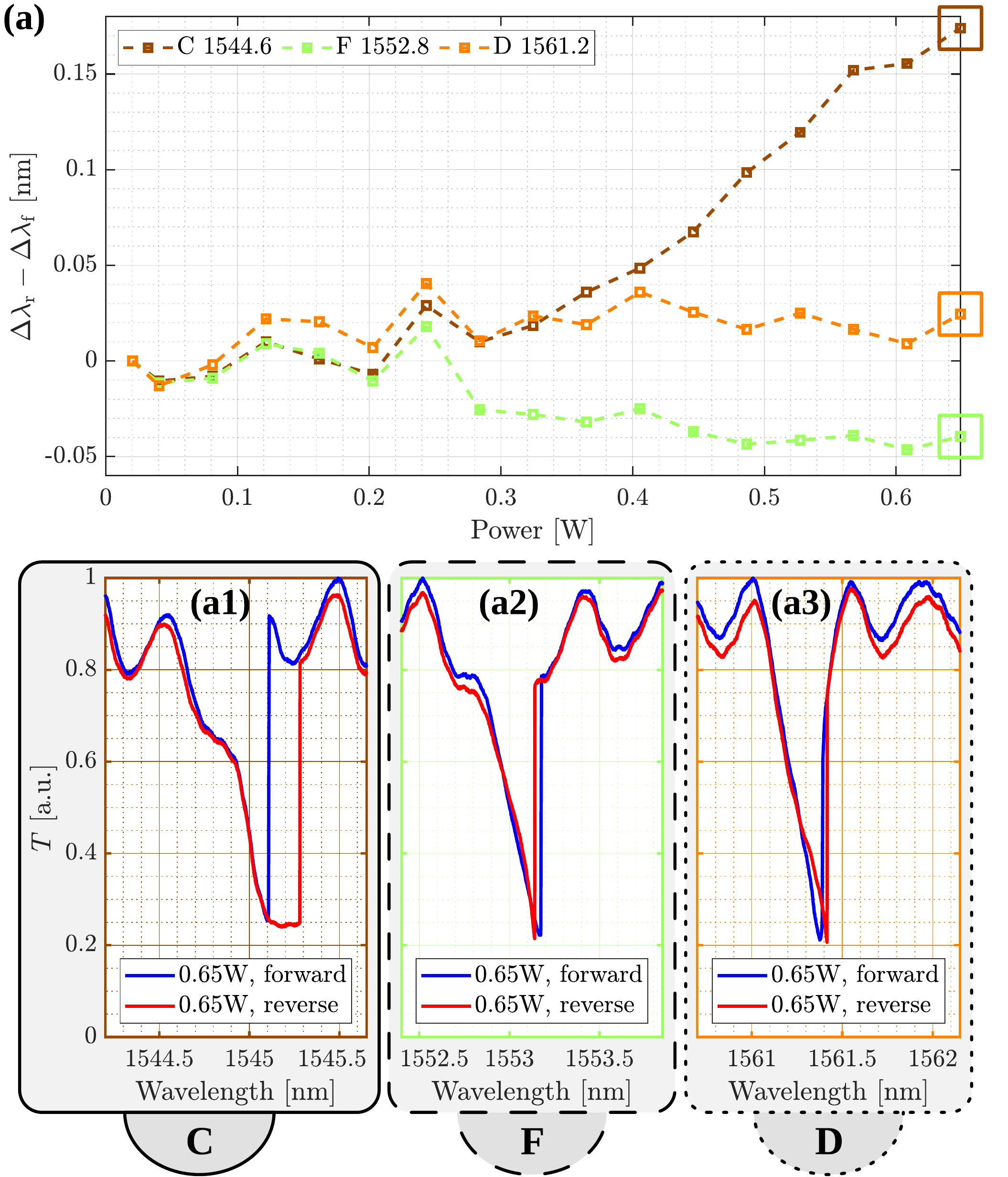}
    \caption{(a) the difference between the resonant shift in the \reverse and \forward configurations as a function of the input power for the three cases described in Fig. \ref{fig:LinearResponse}. The brown, green and orange colors highlight the three different interference cases: \textit{constructive}-like (C), \textit{Fano}-like (F) and \textit{destructive}-like (D) linear regime reflection lineshape. Panels (a1)-(a3) show the normalized transmission spectra at maximum input power for the \forward (blue lines) and \reverse (red lines) configurations and for the three interference cases. In this figure are reported upwards wavelength ramps.} \label{fig:NonLinearMeasu}
\end{figure}

If we look at the experimental results and compare Figs. \ref{fig:NonLinearCostructive} (a) and (b), at sufficiently high $P$, we note that the transmission spectra differ substantially. In particular, there is a wavelength interval where the two transmissions are no longer equal, i.e. where the Lorentz reciprocity is broken \cite{Alberto2021}. We quantify the extent of this wavelength region, by calculating the difference $\Delta \lambda_{\rm r} - \Delta \lambda_{\rm f}(P)$ between the relative shift of $\lambda^{\rm res}$ for the \reverse configuration $\Delta \lambda_{\rm r} (P)= \lambda^{\rm res}_{\rm r} (P)- \lambda^{\rm res}_{\rm r} (P\simeq 0)$ and the \forward configuration $\Delta \lambda_{\rm f} (P)= \lambda^{\rm res}_{\rm f} (P)- \lambda^{\rm res}_{\rm f} (P\simeq 0)$, i.e. between the ``hot'' and ``cold'' resonant wavelengths. $\Delta \lambda_{\rm r} - \Delta \lambda_{\rm f}$ vs $P$ is shown in \figname\ref{fig:NonLinearMeasu} (a) together with representative comparisons between the normalized transmittance spectra at maximum $P$ for the \forward and the \reverse orientations (\figname\ref{fig:NonLinearMeasu} (a1)-(a3)). In \figname\ref{fig:NonLinearMeasu}, the brown, green and orange colors refer to the different wavelength shifts for the \textit{constructive}-like (C), \textit{Fano}-like  (F) and \textit{destructive}-like  (D) linear regime reflection lineshape, respectively.

As already reported in \cite{Alberto2021}, in the \textit{constructive}-like case (blue symbols),  $\Delta \lambda_{\rm r} - \Delta \lambda_{\rm f}(P)$ is positive and, therefore, the \reverse configuration shows a higher resonance shift (see \figname\ref {fig:NonLinearMeasu} (a1)) for all $P$. Since this shift is proportional to the power stored inside the cavity, the \reverse configuration is characterized by a high internal energy. Similarly, the \textit{destructive}-like case shows positive but small  $\Delta \lambda_{\rm r} - \Delta \lambda_{\rm f}(P)$ values (orange symbols and panel (a3) of \figname\ref{fig:NonLinearMeasu}).
On the contrary, the \textit{Fano}-like case exhibits a negative detuning $\Delta \lambda_{\rm r} - \Delta \lambda_{\rm f}(P)$ which implies a higher internal energy in the \forward than in the \reverse configuration (see panel (a2) of \figname\ref{fig:NonLinearMeasu}). This means that reflectance in the \waveguide facets could cancel the effect of the S-shaped waveguide.

\section{Theory and simulation} 
\label{Sec:Theory}
\subsection{Linear regime}
\label{Subsec:LineaRegime}
\begin{figure*}[t!]
    \centering
    \includegraphics[width=1\textwidth]{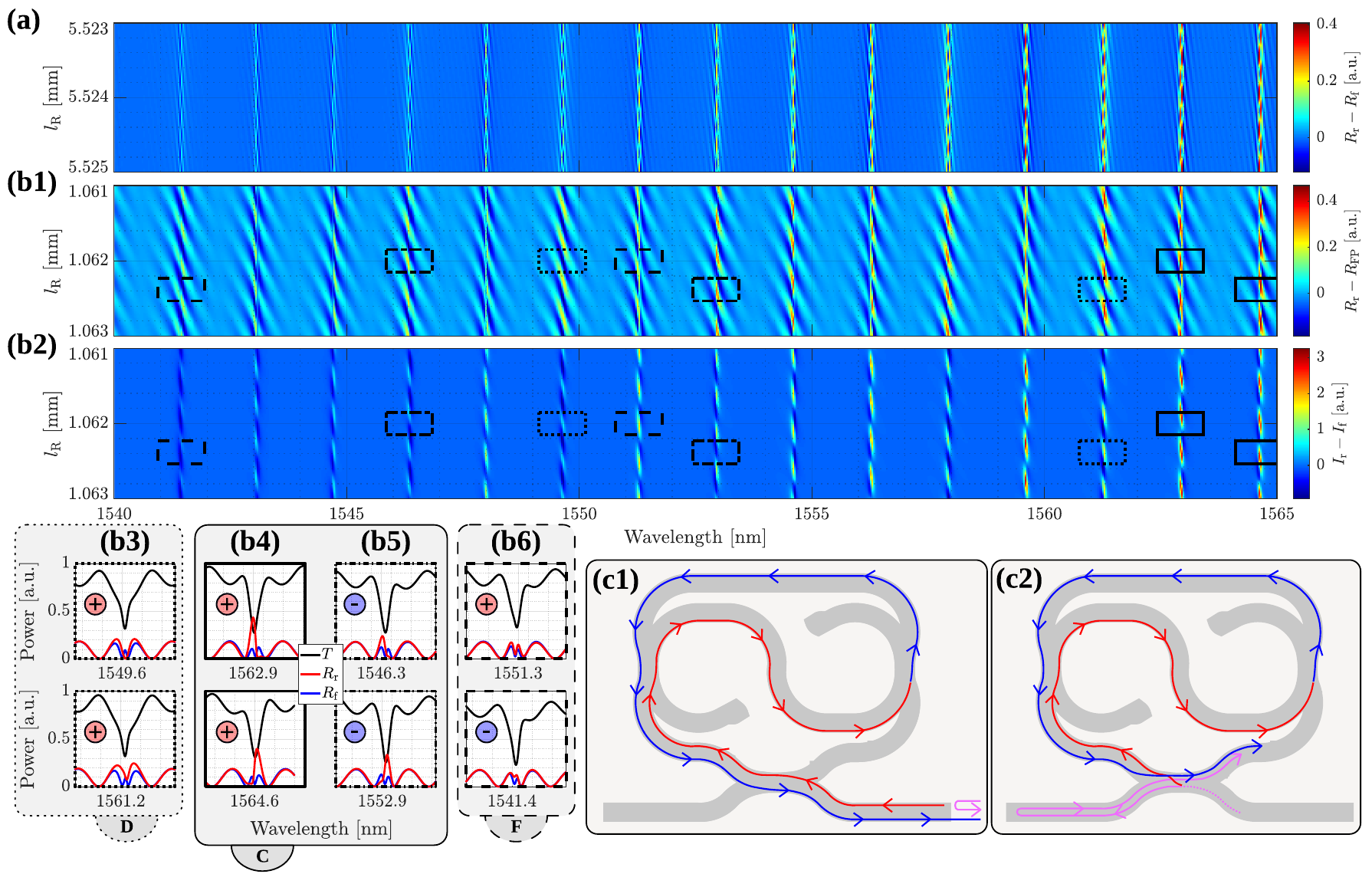}
    \caption{(a) Map of the difference between the reflected intensities for the \forward and \reverse orientation as a function of the incident wavelength ($\lambda$) and the \waveguide length ($l_{\rm R}$). (b1) Map of the difference between the reflected intensities for the \reverse and the \FPO as a function of $\lambda$ and $l_{\rm R}$. (a) and (b1) show a $2\um$-range of $l_{\rm R}$ around $l_{\rm R}=5.524\mm$ and $l_{\rm R}=1.062\mm$, respectively. (b2) Map of the difference between the energy stored within the TJMR in the \reverse ($I_{\rm r}$) and \forward ($I_{\rm f}$) configuration as a function of $\lambda$ and $l_{\rm R} \simeq 1.062 mm$. $l_{\rm L} \simeq 0.431\mm$ is constant in whole maps. In panels (b3), (b4), (b5) and (b6)  are plotted the transmission and reflection spectra for $l_{\rm R}=1.0620\mm$ (top) and $l_{\rm R}=1.0624\mm$ (bottom). The different types of rectangles connects these graphs with the maps (b1) and (b2). specifically, the dotted, dash-dotted/solid and dashed line refer to the \textit{destructive}-like (D), \textit{constructive}-like (C) and \textit{Fano}-like (F) case. The plus and minus signs inside the graphs highlight when the difference of the internal energy between the \reverse and \forward orientation is positive and negative, respectively. Panels (c1) and (c2) show the interference diagrams. The red (blue) arrows labels the CW (CCW) mode.}
    \label{fig:Interference}
\end{figure*}
In order to confirm the role of the \FPOs in the linear and nonlinear regimes, we performed numerical simulations of the device.
These simulations were based on the theoretical model reported in  \cite{Calabrese2020}.
Here, the whole system is modeled by using the transfer matrix method where the only source of back-reflection is given by the \FP cavity generated by the end facets of the \waveguide.
The three directional couplers of \figname\ref{fig:Setup} (a) are schematized by three reciprocal and lossless beamsplitters characterized by their transmission and coupling amplitudes ($t^2+k^2=1$).
The parameters used in these simulations are reported in \figname\ref{fig:Param} of \appname\ref{app:param} and were determined by the geometry of the device and by a fit of the transmission spectrum of \figname\ref{fig:LinearResponse} (b). A wavelength-dependent effective refractive index as in \cite{Calabrese2020} was also used. Note that, for these simulations $l_{\rm L}=0.431\mm$ is fixed.

As we are interested in understanding the role of the \FP fringes, we computed the device reflectivity $R$ for the \reverse ($R_{\rm r}$) and the \forward ($R_{\rm f}$) configurations and for the case without coupling between the \waveguide and the TJMR ($t_{1}=1$, as defined in Fig. \ref{fig:Setup} (a)). This last quantity describes the contribution to the reflectance of the device due to the \FP in the \waveguide and is labeled $R_{\rm FP}$. \fignametot\ref{fig:Interference} (a) and (b1) show the  $\lambda$ vs $l_{\rm R}$ maps of $R_{\rm r}-R_{\rm f}$ and of $R_{\rm r}-R_{\rm FP}$. A $2\um$-range of $l_{\rm R}$ around a value of $l_{\rm R}=5.524\mm$ (Fig. \ref{fig:Interference} (a)) and $l_{\rm R}=1.062\mm$ (Fig. \ref{fig:Interference} (b1)) is mapped. Since interference effects affect the internal energy ($I$) in the TJMR, we also plot in Fig. \ref{fig:Interference} (b2) the $\lambda$ vs $l_{\rm R}$ map of the difference between the internal energies in the \reverse ($I_{\rm r}$) and \forward orientations ($I_{\rm f}$) for the short device case. More details on the calculation of $I_{\rm r}$ and $I_{\rm f}$ are reported in \appname\ref{app:energy}.
These various differences show the unidirectional behavior of the device and the role of the \waveguide in this phenomenon. In particular, the clear lines that cut vertically through the maps represent the TJMR resonances.
The colors reflect the different values of $R_{\rm r}-R_{\rm f}/R_{\rm r}-R_{\rm FP}/I_{\rm r}-I_{\rm f}$ for each resonance. These take into account the spectral dispersion of the effective refractive index, of the propagation losses, of the coupling parameters and the $l_{\rm R}$ dependence of the interference.
For long $l_{\rm R}$ (\figname\ref{fig:Interference} (a)) the fact that $R_{\rm r}-R_{\rm f}$ always shows clear peaks is in agreement with the experimental data of \figname\ref{fig:LinearResponse} (a). For short $l_{\rm R}$ (\figname\ref{fig:Interference} (b1)), the decrease of the \waveguide length allows catching all the experimental cases. These are highlighted by the rectangles in \figname\ref{fig:Interference} (b1)-(b2). Specific examples of the simulated transmission and reflection lineshapes for the \textit{destructive}-like (D), \textit{constructive}-like (C) and \textit{Fano}-like (F) cases are shown in \figname\ref{fig:Interference} (b3), b(4)-b(5), and (b6), respectively, for $l_{\rm R}=1.0620\mm$ (top) and $l_{\rm R}=1.0624\mm$ (bottom).

Let us start from the \textit{destructive}-like case. This is characterized by a dip of the reflectance in the \reverse configuration (\figname\ref{fig:LinearResponse} (b3)). This case is exemplified by the dotted rectangles in panels (b1) and (b2) and by the lineshapes in (b3) of \figname\ref{fig:Interference}. The reflectance dip is a consequence of the interference between the light that is reflected at the input facet of the \waveguide (magenta arrow) and the light that propagating in the CW mode (red arrows) is coupled into the CCW one through the S waveguide (blue arrows), as shown in the sketch of panel (c1).  When such interference is destructive, the reflected intensity can exhibit a dip. The condition for interference in the device is:
\begin{equation}
\frac{2\pi}{\lambda}n_{\rm eff}(2l_{\rm R}+2z_3+z_4+z_2) +2\pi = \pi + \pi +2\pi m_{\rm I1}\,,
\label{equ:IntDistFacFabry}
\end{equation}
where $n_{\rm eff}$ and $m_{\rm I1}$ are, respectively, the effective refractive index and an integer number.


Thus, the phase difference between the path followed by the light reflected from the TJMR (left hand of \eqname\eqref{equ:IntDistFacFabry}) and the one followed by the light reflected from the input facet must be an odd multiple of $\pi$. That is, when $m_{\rm I1}$ satisfies:
\begin{equation}
m_{\rm I1} = \frac{n_{\rm eff}(2l_{\rm R}+2z_3+z_4+z_2)}{\lambda}\,.
\end{equation}
This condition is satisfied in the example shown in panel (b3, top), where the reflectance (red line) reduces to zero at the resonant wavelength. However, as shown in panel (b3, bottom), a slight shift of the \FP fringes due to a slight variation in $l_{\rm R}$ (from 1.0620$\mm$ (top) to $1.0624\mm$ (bottom)) causes a different interference which yields a non-zero reflection. This interference pattern is also confirmed by the positive value of the internal energy difference shown in panel (b2), as evidenced by the dotted rectangle. Hence, less reflection of the device does not mean less internal energy in the \reverse with respect to the \forward configurations. This can be understood by considering two other interference diagrams.
The first one is defined by the path followed by the light in the \waveguide. It gives rise to the typical constructive \FP interference at the exit of the input facet:
\begin{equation}
\frac{2\pi}{\lambda}n_{\rm eff}(2l_{\rm R}+2l_{\rm L}) = \pi +2\pi m_{\rm FPCR}\,,
\end{equation}
where $m_{\rm FPCR}$ is an integer number.  The second is more complex and is shown in panel (c2) of \figname\ref{fig:Interference}. It is given by the constructive interference, inside the TJMR, between the light which is transferred from the S-shaped waveguide to the CCW mode (from red to blue arrows) and the one that is reflected from the output facet of the \waveguide (magenta arrows).  Defining $m_{\rm I2}$ as an integer number, this interference occurs when the following relation is satisfied:
\begin{equation}
\frac{2\pi}{\lambda}n_{\rm eff}2l_{\rm L} +\frac{\pi}{2} = \frac{\pi}{2} + \frac{2\pi}{\lambda}n_{\rm eff}(2z_3+z_4+z_2) + \pi +2\pi m_{\rm I2}\,.
\end{equation}
These three numbers $m_{\rm FPCR}$, $m_{\rm I1}$ and $m_{\rm I2}$ are strictly connected as: 
\begin{equation}
m_{\rm FPCR} = m_{\rm I1} + m_{\rm I2}\,.
\label{eqa:integer}
\end{equation}
If $m_{\rm FPCR}$ and $m_{\rm I1}$ are integer numbers, then also $m_{\rm I2}$ is an integer number. In other words, if the \FP interference exhibits a peak and the device reflection shows a dip, then inside the TJMR occurs a constructive interference with a build up of internal energy ($I_{\rm r}-I_t>0$).

\begin{figure*}[t!]
    \centering
    \includegraphics[width=1\textwidth]{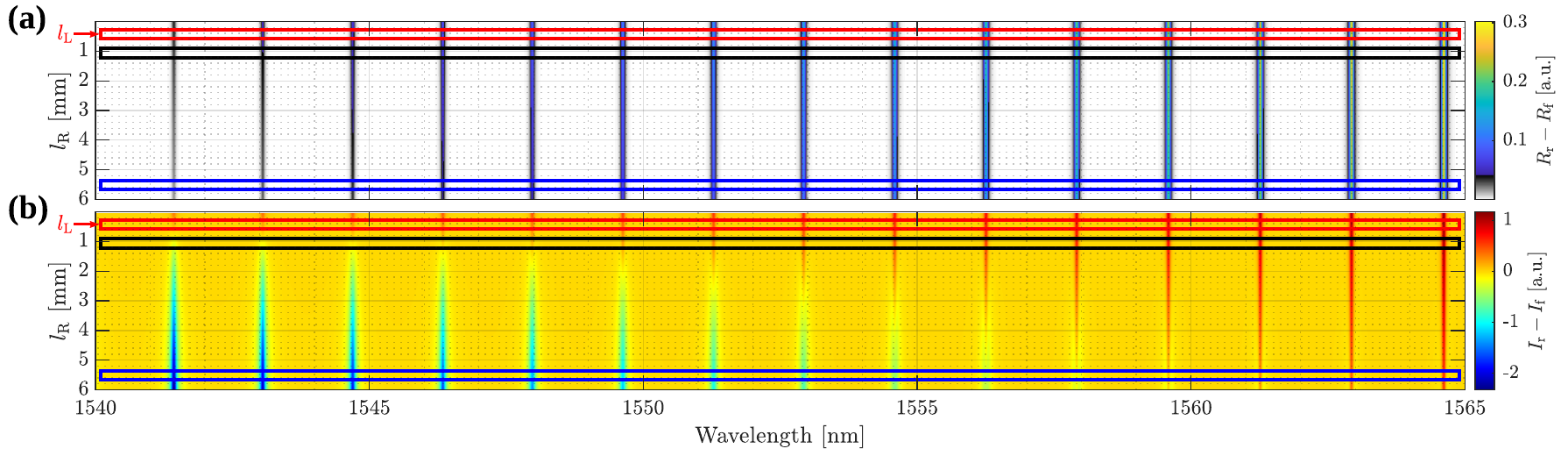}
    \caption{(a) Map of difference of the reflected intensities in the \reverse and \forward configurations as a function of the \waveguide length $l_{\rm R}$ and of the wavelength $\lambda$. (b) $\lambda$ vs $l_{\rm R}$ map of the difference of the microresonator internal energy between the \reverse and \forward orientations. In these maps, $l_{\rm L} \simeq 0.431\mm$ is constant. The red, black and blue rectangles highlight, respectively, the regions where the $l_{\rm R}=l_{\rm L}=0.431\mm$, $l_{\rm R}=1.062\mm$ and $l_{\rm R}=5.52\mm$.}
    \label{fig:maps_NoFP}
\end{figure*}

This analytical model also explains the \textit{constructive}-like case shown in \figname\ref{fig:LinearResponse} (b1). The solid and dashed-dotted rectangles highlight regions characterized by a high reflection intensity (\figname\ref{fig:Interference} (b1)) but different internal energies (\figname\ref{fig:Interference} (b2)). Characteristic spectra are plotted in panel (b4) and (b5) for $l_{\rm R}=1.0620\mm$ (top) and $l_{\rm R}=1.0624\mm$ (bottom). In this case, the TJMR behaves as a unidirectional reflector. Therefore, the reflected intensity exhibits a maximum in the \reverse configuration (red lines).
The panel (b4) differs from panel (b5) because of the difference between the internal energy of the \forward and \reverse configurations. In the first, the stored energy is higher in the \reverse orientation than in the \forward one. In the second, a lower energy is found in the \reverse than in the \forward configuration. The difference between the two situations is due to the wavelength dependence of the propagation losses in the \waveguide (\appname\ref{app:param}), as we will discuss in the following.

Panel (b1) of \figname\ref{fig:Interference} shows also the \textit{Fano}-like case (see dashed rectangles) as highlighted by the graphs of panel (b6). This is an intermediate case between the \textit{constructive}-like and the \textit{destructive}-like cases.


The TJMR loses its fundamental property of being a unidirectional reflector because of the \FPOs.
Simulating the response of the device in the absence of the \FP (i.e with zero facets reflectivity), we obtain the $\lambda$ vs $l_{\rm R}$ maps in \figname\ref{fig:maps_NoFP}. Note that, for these simulations $l_{\rm L}=0.431\mm$ is fixed. In particular, \figname\ref{fig:maps_NoFP} (a) shows $R_{\rm r}-R_{\rm f}$ while \figname\ref{fig:maps_NoFP} (b) shows $I_{\rm r}-I_{\rm f}$. The red, black, and blue rectangles highlight the regions around the values $l_{\rm R}=l_{\rm L}=0.431\mm$, $l_{\rm R}=1.062\mm$, and $l_{\rm R}=5.52\mm$, respectively, i.e. in the latter two the TJMR is not placed in a symmetric position with respect to the two \waveguide facets. In contrast with \figname\ref{fig:Interference}, no oscillations are observed and at the resonances $R_{\rm r}-R_{\rm f}>0$ always since $R_{\rm f}=0$. Note that $R_{\rm r}$ changes as $l_{\rm R}$ varies. In fact, as $l_{\rm R}$ increases, the \waveguide propagation losses affect the amount of light coupled to the microresonator. Therefore, less energy is transferred from the CW to the CCW mode. As a function of $l_{\rm R}$ (see the rectangles), the reflected intensity in the \reverse configuration increases with the wavelength. This is due to the spectral dependence of the \waveguide propagation losses, which are large at $1540\nm$ and decrease monotonically as $\lambda$ increases (see \appname\ref{app:param}). Also $I_{\rm r}-I_{\rm f}$ is affected by the relative position of the TJMR with respect to the \waveguide. Indeed, by increasing $l_{\rm R}$ more and more resonances present a negative $I_{\rm r}-I_{\rm f}$. Moreover, this negative value becomes larger as $\lambda$ decreases, i.e. as the losses increase.

\begin{figure*}[t!]
    \centering
    \includegraphics[width=1\textwidth]{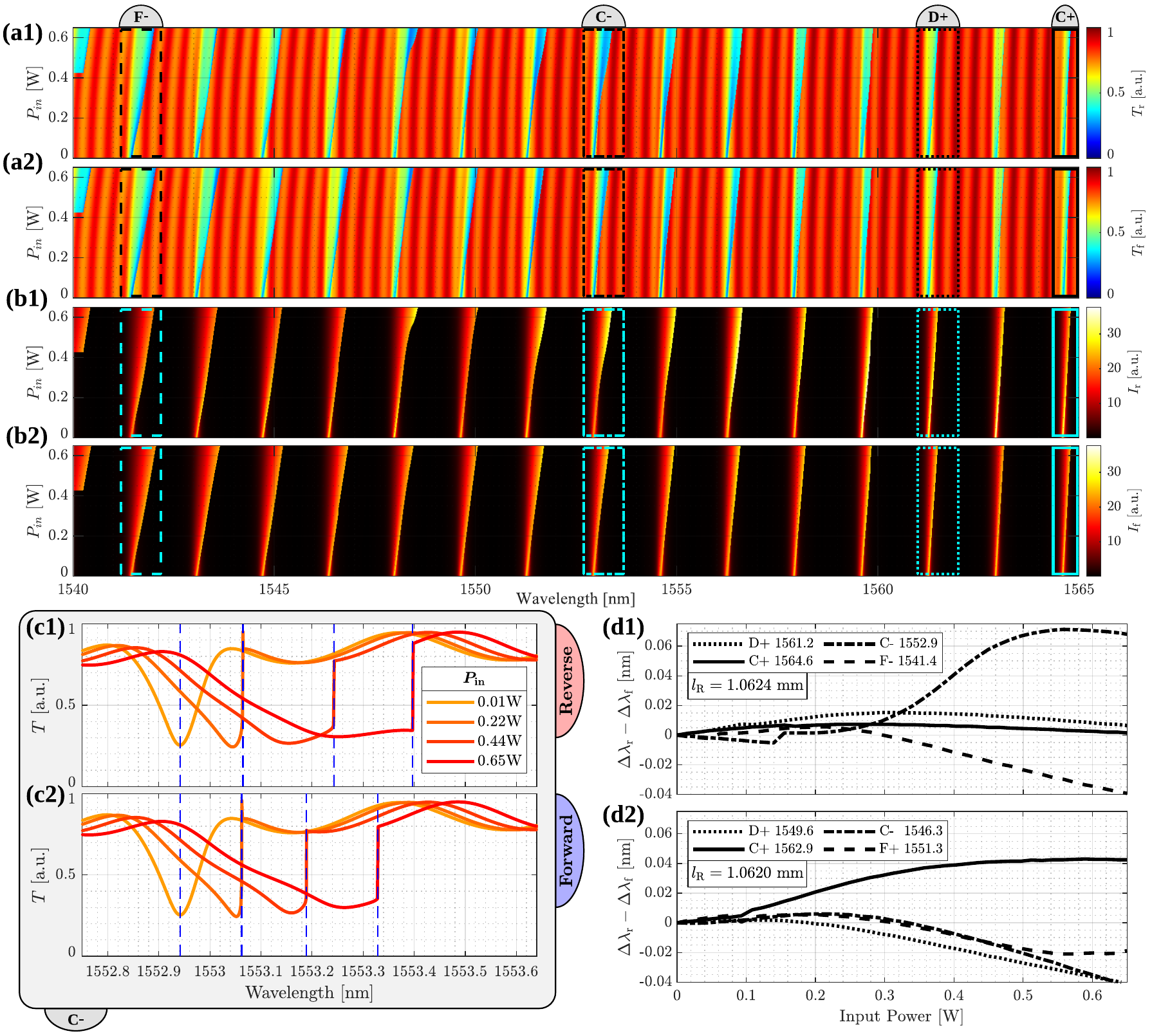}
    \caption{(a1) Spectral transmission ($T_{\rm r}$) map as a function of the input power ($P_{\rm in}$) for the \textit{reverse} configuration. (a2)  $P_{\rm in}$ vs $\lambda$ map of the transmission ($T_{\rm f}$) for the \textit{forward} configuration. (b1) $P_{\rm in}$ vs $\lambda$ map of the internal energy ($I_{\rm r}$) of the TJMR for the \textit{reverse} configuration. (b2) $P_{\rm in}$ vs $\lambda$ map of the internal energy ($I_{\rm f}$) of the TJMR for the \textit{forward} configuration. These maps are simulated by assuming an increasing $\lambda$ scan. (c1)-(c2) Transmittance for the C- case, i.e. \textit{constructive}-like shape where $I_{\rm r}-I_{\rm f}<0$ in the linear regime, for different input powers. The wavelength is scanned from low to high values. Precisely, (c1) and (c2) show the \reverse and \forward orientations, respectively. (d1) Difference between the resonances shift in the \reverse and \forward configurations, $\Delta\lambda_{\rm r}-\Delta\lambda_{\rm f}$, as a function $P_{\rm in}$. For (a1), (a2), (b1), (b3), (c1), (c2) and (d1) the \waveguide length is $l_{\rm R} = 1.0624\mm$. (d2)  $\Delta\lambda_{\rm r}-\Delta\lambda_{\rm f}$ vs $P_{\rm in}$ for $l_{\rm R} =1.0620\mm$. The dashed, dotted-dashed, dotted and solid rectangles allow relating the maps (a1)-(a2) and (b1)-(b2) to the graph in panel (d1). The letters D, C and F denotes, in the linear regime, the \textit{destructive}-like, \textit{constructive}-like and \textit{Fano}-like cases, respectively. The plus and minus signs label a positive and negative difference between the internal energies of the \reverse and \forward configurations in the linear regime.   
    }
    \label{fig:NL_maps}
\end{figure*}

To summarize the analysis of the linear regime, the interference between the reflected fields at the ends of the \waveguide and by the TJMR generate different spectral responses. In particular, depending on the period of the \FP fringes, the device may preserve or lose its unidirectional reflector nature. As a result, the difference between the internal energies in the \reverse and \forward configurations may assume both positive and negative values.       

\subsection{Nonlinear regime}

The device is modelled in the nonlinear regime by following the finite-element model developed in Ref.~\cite{Alberto2021}. The light propagation inside the device is obtained by solving the nonlinear Helmholtz equation while taking also into account reflection at the \waveguide facets. We took the thermal nonlinearity parameters from \cite{Trenti_2018}. The set of employed parameters is shown in \appname\ref{app:param} and \ref{app:model}.

\fignametot\ref{fig:NL_maps} (a) and (b) show the transmission spectra and the TJMR internal energies as a function of the input power ($P_{\rm in}$) for $l_{\rm R} = 1.0624\mm$, while scanning $\lambda$ from low to high values. Panels (a1) and (b1) display the \reverse configuration while (a2) and (b2) show the \forward one.  As expected, increasing $P_{\rm in}$, the resonances shift proportionally to the internal energy due to the nonlinear refractive index. This shift is towards longer $\lambda$ in agreement with the positive sign of the nonlinear coefficient (see \figname\ref{fig:nT} in \appname\ref{app:model}). Notice that the \FP fringes slightly shift to longer wavelengths too. The difference between the resonance and fringe shifts is explained by the larger energy stored in the microresonator than in the \waveguide. In fact, a $9$ times field enhancement factor is computed for the TJMR.
Within the maps, we can identify the different features seen in the experimental section, i.e. \textit{constructive}-like (C), \textit{destructive}-like (D) and \textit{Fano}-like (F) shape. These are labelled with a + (-) when, in the linear regime, $I_{\rm r} - I_{\rm f} > 0$ ($I_{\rm r} - I_{\rm f} < 0$).  \fignametot\ref{fig:NL_maps} (c1) and (c2) are the theoretical analogue of \fignametot\ref{fig:NonLinearCostructive} (b) and (a), which show the experimental transmission spectra. \fignametot\ref{fig:NL_maps} (c1) and (c2) display the transmittance for different input powers in the C- case. The wavelength is scanned from low to high values. In particular, panel (c1) and (c2) show the \reverse and \forward orientations, respectively. The theoretical model reproduces the experimental behavior and Lorentz reciprocity breaking appears through a different resonance shift between forward and reverse orientation increasing the input power.
Comparing the nonlinear shift for the \forward and \reverse orientations, we do not observe a regular trend. \figname\ref{fig:NL_maps} (d1) shows $\Delta\lambda_{\rm r}-\Delta\lambda_{\rm f}$ as a function of $P_{\rm in}$, computed from the (a1)-(a2) maps. Specifically, the dotted, dash-dotted/solid and dashed lines highlight the \textit{destructive}-like (D), \textit{constructive}-like (C) and \textit{Fano}-like (F) cases, respectively. These resonances are the ones shown in \figname\ref{fig:Interference} (b3)-(b6) for the linear regime and $l_{\rm R} \simeq 1.0624\mm$ (i.e. the bottom panels). It is observed that $\Delta\lambda_{\rm r}-\Delta\lambda_{\rm f}$ shows different behaviors in the three cases in agreement with the experimental results of \figname\ref{fig:NonLinearCostructive}. In fact, in both the experimental (labeled D in \figname\ref{fig:NonLinearMeasu}) and the theoretical case (labeled D+ in \figname\ref{fig:NL_maps}), the \textit{destructive}-like case shows a positive value of $\Delta \lambda_{\rm r}-\Delta \lambda_{\rm f}$ slightly greater than zero. The same agreement holds for the experimental (C) and theoretical (C-) \textit{constructive}-like cases where the detuning is always positive and reaches a maximum value around $0.07\nm$. Similarly for the \textit{Fano}-like case, where both the theoretical (F-) and experimental (F) shift differences show negative values.

However, a clear relation between $I_{\rm r}-I_{\rm f}$ in the linear and nonlinear regimes does not emerge.
In fact, in the \textit{constructive} case, the $\Delta\lambda_{\rm r}-\Delta\lambda_{\rm f}$ vs $P_{\rm in}$ curve shows both a positive slope for the C- situation, where  $I_{\rm r}-I_{\rm f}<0$ in the linear regime, as well as an almost zero slope for the C+ situation where  $I_{\rm r}-I_{\rm f}>0$ in the linear regime. This lack of a direct relation between $I_{\rm r}-I_{\rm f}$ in linear and nonlinear regimes is also shown in \figname\ref{fig:NL_maps} (d2). It displays $\Delta\lambda_{\rm r}-\Delta\lambda_{\rm f}$ vs $P_{\rm in}$ for the resonances shown on the top panels (b3)-(b6) of \figname\ref{fig:Interference} (i.e. when $l_{\rm R} \simeq 1.0620 \mm$).
Here, the \textit{destructive}-like case (D+) presents a negative $\Delta\lambda_{\rm r}-\Delta\lambda_{\rm f}$ shift despite $I_{\rm r}-I_{\rm f}>0$ in the linear regime. In addition, the \textit{constructive}-like case with $I_{\rm r}-I_{\rm f}<0$ (C-) exhibits a negative $\Delta\lambda_{\rm r}-\Delta\lambda_{\rm f}$ shift in contrast with \figname\ref{fig:NL_maps} (d1). Therefore, depending on their spectral position, the resonances of the TJMR show a different $\Delta\lambda_{\rm r}-\Delta\lambda_{\rm f}$ shift which we attribute to the interplay between the \FP and the asymmetric losses ($l_{\rm L} \neq l_{\rm R}$) in the \waveguide.

\begin{figure}[t!]
    \centering
    \includegraphics[width=0.45\textwidth]{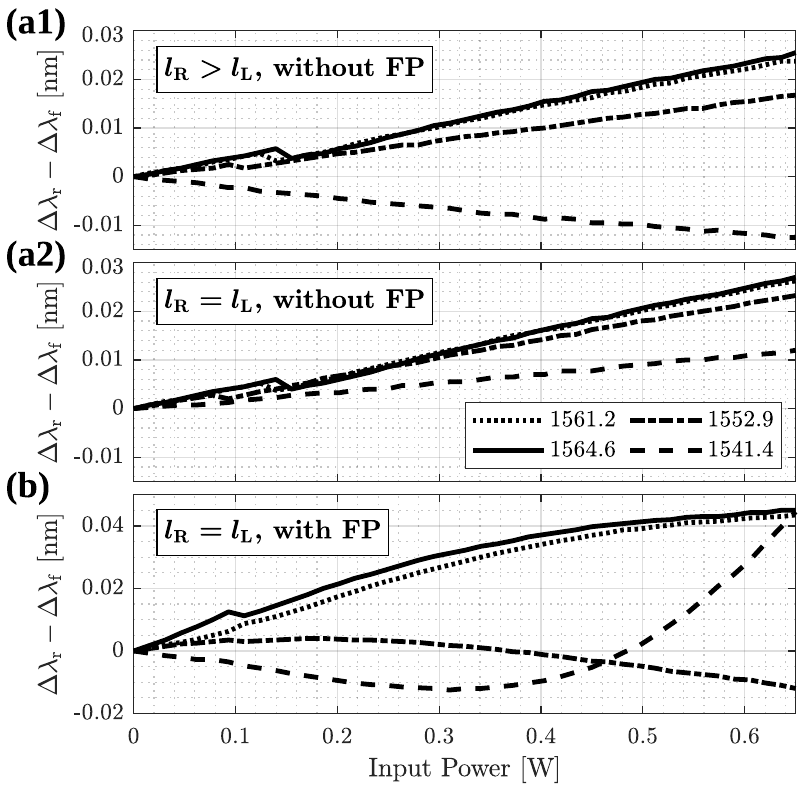}
    \caption{Difference of the resonance shift between the \reverse and \forward configurations ($\Delta\lambda_{\rm r}-\Delta\lambda_{\rm f}$) as a function of the input power. (a1) and (a2) show the simulation results neglecting the \FPOs for a device where the TJMR is placed in an asymmetric or a symmetric position with respect to the \waveguide facets, respectively. $l_{\rm R}$ and $l_{\rm L}$ are the \waveguide lengths defined in \figname\ref{fig:Setup}. In (a1) $l_{\rm R}  = 1.0624 \mm$ and $l_{\rm L} = 0.431 \mm$. In (a2) $l_{\rm R} =l_{\rm L} = 0.431 \mm$. (b) $\Delta\lambda_{\rm r}-\Delta\lambda_{\rm f}$ computed taking into account the \FP effect in the \waveguide for $l_{\rm R} =l_{\rm L} = 0.431 \mm$. }
    \label{fig:NL_Delta_lam_r-f}
\end{figure}

This is evidenced in \figname\ref{fig:NL_Delta_lam_r-f}. When the \FP effect is switched off by zeroing the reflection coefficients at the \waveguide facets, $\Delta \lambda_{\rm r}-\Delta \lambda_{\rm f}$ grows linearly with $P_{\rm in}$. The different slopes are related to the values of the \waveguide propagation losses. Negative and positive slope values are due to larger or smaller asymmetric losses. In fact, the maximum slope appears at longer wavelengths where the losses are smaller (see dashed and solid lines for $1564.6 \nm$ and $1561.2 \nm$ in \figname\ref{fig:NL_Delta_lam_r-f} (a1)). When the losses are symmetric, i.e the TJMR is placed in a symmetric position, the $\Delta \lambda_{\rm r}-\Delta \lambda_{\rm f}$ slopes are always positive (\figname\ref{fig:NL_Delta_lam_r-f} (a2)). Furthermore, when the \FP effect is switched on, a more complicated scenario appears (\figname\ref{fig:NL_Delta_lam_r-f} (b)). The $\Delta \lambda_{\rm r}-\Delta \lambda_{\rm f}$ does no longer show a linear $P_{\rm in}$ dependence and negative or positive values appear even when the losses are symmetric. Here, $\Delta \lambda_{\rm r}-\Delta \lambda_{\rm f}$ shows variations strictly connected to the interference between the fields reflected by the end facets of the \waveguide and the one reflected within the TJMR. The phase relation between these fields is given by the different variations of the nonlinear refractive index inside the TJMR and the \waveguide. Interestingly, as shown in \figname\ref{fig:NL_Delta_lam_r-f} (b), even with a symmetric \waveguide the \FP fringes could drastically change the shift of the resonances. As a result, a positive (negative) difference of the resonance shift may become negative (positive) by increasing the input power (see dotted-dashed and dashed line in \figname\ref{fig:NL_Delta_lam_r-f} (b)). Therefore, we can conclude that the combined action of the \FP and of the asymmetric losses due to the \waveguide can compensate the effect of the S-shaped waveguide in the TJMR leading to a higher internal energy in the \forward configuration than in the \reverse configuration. In fact, since $l_{\rm R}>l_{\rm L}$, more light attenuation is observed in the \reverse than in the \forward configuration. 
It is worth noticing that the presence of the \FP effect increases the wavelength interval $\Delta \lambda_{\rm r}-\Delta \lambda_{\rm f}$ where the Lorentz reciprocity is broken. This is observed by comparing \figname\ref{fig:NL_maps} (d1) and \figname\ref{fig:NL_Delta_lam_r-f} (a1). In the first, $\Delta \lambda_{\rm r}-\Delta \lambda_{\rm f} \simeq$ $0.07\nm$ while in the second, $\Delta \lambda_{\rm r}-\Delta \lambda_{\rm f}<$ $0.03\nm$.

\section{Conclusion}
\label{Sec:Conc}

In this work, we have theoretically and experimentally shown how the  properties of the bus waveguide influence the linear and nonlinear response of the taiji microresonator. Indeed, both the Fabry-Perot effect, due to the bus waveguide end facets reflection, and the asymmetric propagation losses along the bus waveguide affect the measured and simulated responses. In the linear regime, the experimental spectra are well explained by an analytical model based on the transfer matrix method and by using intuitive interference diagrams. Increasing the period of the Fabry-Perot oscillations, the device does not preserve its functionality as a unidirectional reflector. Indeed, the interference between the reflected field at the input facet of the bus waveguide and the one reflected within the taiji can also reduce the device reflectivity to zero.

Furthermore, the Fabry-Perot can redistribute the taiji microresonator internal energy between the clockwise and counterclockwise modes and, thus, strongly modify the nonlinear response. In this nonlinear regime, the different powers stored inside the taiji microresonator are the base of the Lorentz reciprocity breaking in the device.  The breaking appears as a distinct difference between the resonance shifts in the \reverse and \forward configuration. Depending on the specific configuration, the Fabry-Perot effect in the bus waveguide can either reduce or increase the wavelength region where the Lorentz reciprocity breaking is observed. Using a numerical finite-element model we have explained the experimental observations in terms of a different shift between the resonant wavelengths and the Fabry-Perot fringes. Moreover, we demonstrated that a critical role is also played by the propagation losses in the bus waveguide. Indeed, when the taiji microresonator is placed in an asymmetric position with respect to the bus waveguide ends, a variation in the taiji microresonator internal energy also stems from the interplay between the asymmetric propagation losses and the field enhancement due to the microresonator. However, this asymmetry does not influence the unidirectional behavior of the taiji microresonator at sufficiently low input power, i.e. in the linear regime.

Finally, let us note that the device we studied here can be understood as a sophisticated example of a pair of coupled  resonators. Therefore, this work is a starting point towards the study of more complex structures, where an active control of the feedback between nonlinear resonators is used. This allows controlling the violation of the Lorentz reciprocity, and therefore, holds interesting promise for exploiting nonlinear non-Hermitian physics in integrated devices.

\appendix
\subsection{Appendix: parameters of the simulations}
\label{app:param}

\begin{figure}[b!]
    \centering
    \includegraphics[width=0.45\textwidth]{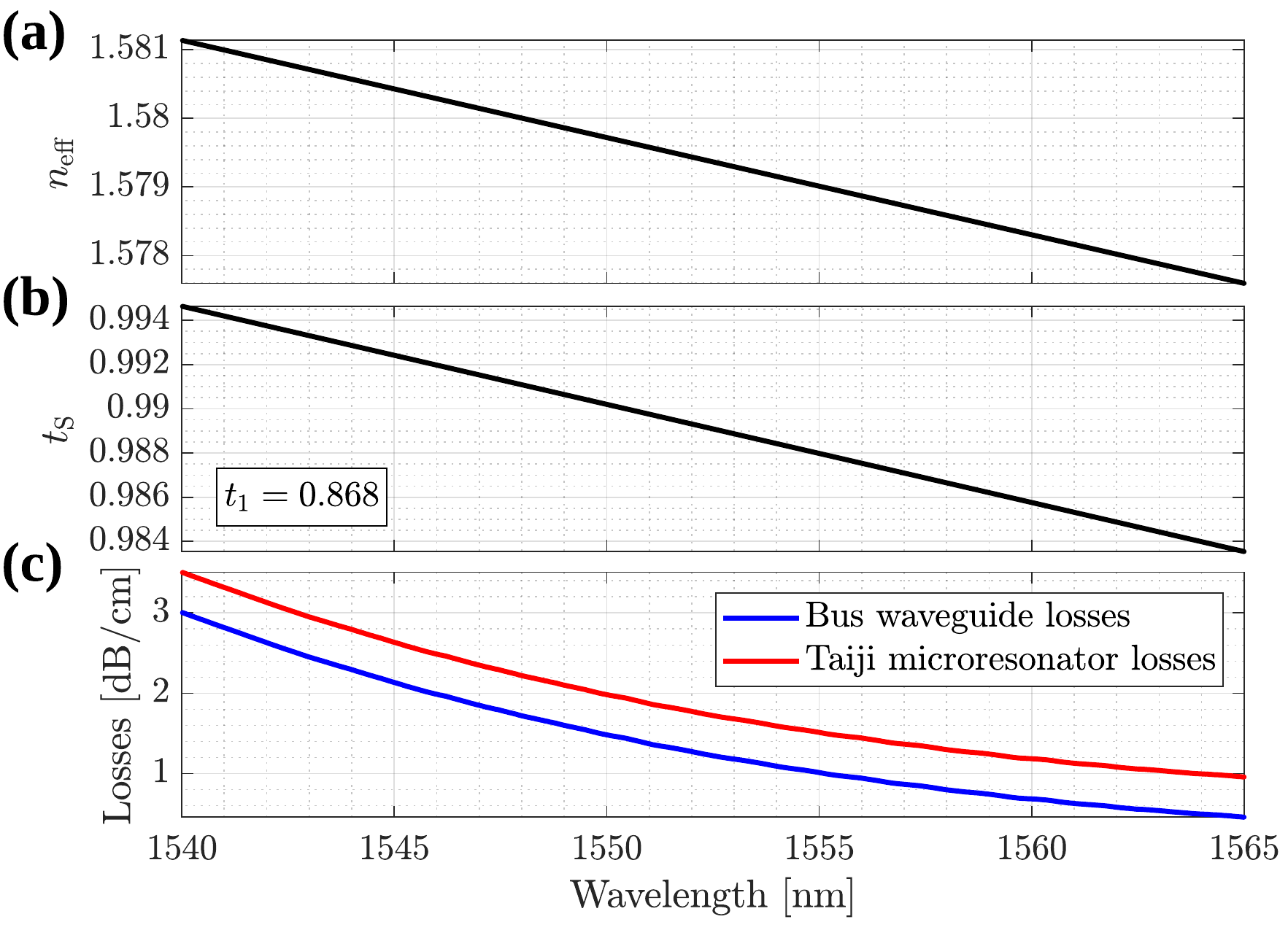}
    \caption{(a) Spectral dependence of the effective mode index. (b) Spectral dependence of the transmission coefficient $t_{\rm S}$. (c) Spectral dependence of the propagation losses for the \waveguide (blue line) and the microresonator (red line).}
    \label{fig:Param}
\end{figure}

In order to model the experiments, we set the parameters of the device as follow.
The perimeter of the taiji racetrack microresonator is fixed imposing $z_1\simeq206\um$, $z_2\simeq398\um$, $z_3\simeq206\um$, while the S-shaped waveguide length is $z_4\simeq391\um$. All these values were derived from the design of the TJMR.
The effective mode index was extrapolated by slightly modifying the one reported in \cite{Calabrese2020} to match the taiji experimental resonances (see \figname\ref{fig:Param} (a)).
The \waveguide length $l_{\rm L}$ was measured from the design $l_{\rm L} \simeq 0.431\mm$. $l_{\rm R}\simeq1.062\mm$ and the reflection coefficients ($r_{\rm L, \,R} = 0.23$) were extrapolated from the experimental \FP fringes.
The spectral dependence of the transmission coefficients $t_2=t_3=t_{\rm S}$,  and of the losses  were estimated by measuring the transmittance, the reflectance and the propagation losses (see \figname\ref{fig:Param}). $t_1=0.868$. By fitting the experimental spectra in the linear and nonlinear regimes, we observed lower propagation losses in the \waveguide than in the TJMR. This difference is due to the bending loss in the microresonator.

\subsection{Appendix: taiji microresonator internal energy calculation}
\label{app:energy}

To simulate the device in the nonlinear regime, it is needed to evaluate the internal energy in the following regions: microresonator, S-shaped waveguide and \waveguide. Since the method is the same, we will describe only the calculation of the microresonator internal energy. Following \cite{Calabrese2020}, the TJMR can be analyzed through twelve different electric fields. Precisely, half of them fields propagates in the CW direction and the other half in the CCW one. All of these fields can be computed by solving the system of equations shown in \cite{Calabrese2020}, for the linear regime or by iterating until convergence for the nonlinear one. To determine the internal energy it is first necessary to calculate the CW and CCW fields at each point of the microresonator. We start from the CW direction and use the fact that from one coupling region to the next and along the wave propagation direction the electric field can be described as $E(z) = E(z_0) e^{i\gamma (z-z_0)}$, where $z_0$ is the starting position, $z$ is the coordinate along the waveguide, and $\gamma$ is a complex parameter that accounts for both phase variation and propagation losses ($\gamma = \frac{2\pi}{\lambda}n_{\rm eff}+i\alpha$). By transfer matrix multiplication, we compute all the CW ($E^{\rm CW}$) and CCW ($E^{\rm CCW}$) fields. Then, the internal energy is the integral of $\left|E^{\rm CCW}+E^{\rm CW}\right|^2$ along the microresonator.

\subsection{Appendix: simulation model}
\label{app:model}

In the linear regime, we used the model presented in \cite{Calabrese2020} to simulate the device. In the nonlinear regime, we extended the equations taking into account that the refractive index is not only wavelength dependent but varies also as a function of the intensity of the electromagnetic wave.
As seen in \cite{Alberto2021}, $n_{\rm eff} = n_{\rm L}+n_{\rm T}I_{\rm thermal}+n_{\rm K}\left(|E^{\rm CCW,\,CW}|^2+2|E^{\rm CW,\,CCW}|^2\right)$, where $n_{\rm L}$ is the refractive index in the linear regime, $n_{\rm T}$ is the coefficient of the thermo-optic nonlinearity, $n_{\rm K}=8\times10^{-16}\cmqW\ll n_{\rm T}$ is the Kerr nonlinear index and $I_{\rm thermal}$ is the total electromagnetic intensity for the three different regions: microresonator, S-shaped waveguide and \waveguide.
To obtain transmissions, reflections, and internal energies as a function of wavelength we process the spectra of all electric fields within the system starting at shorter wavelengths and for each wavelength we evolve the system of field equations to their convergence.

\begin{figure}[ht]
	\centering
	\includegraphics[width=0.45\textwidth]{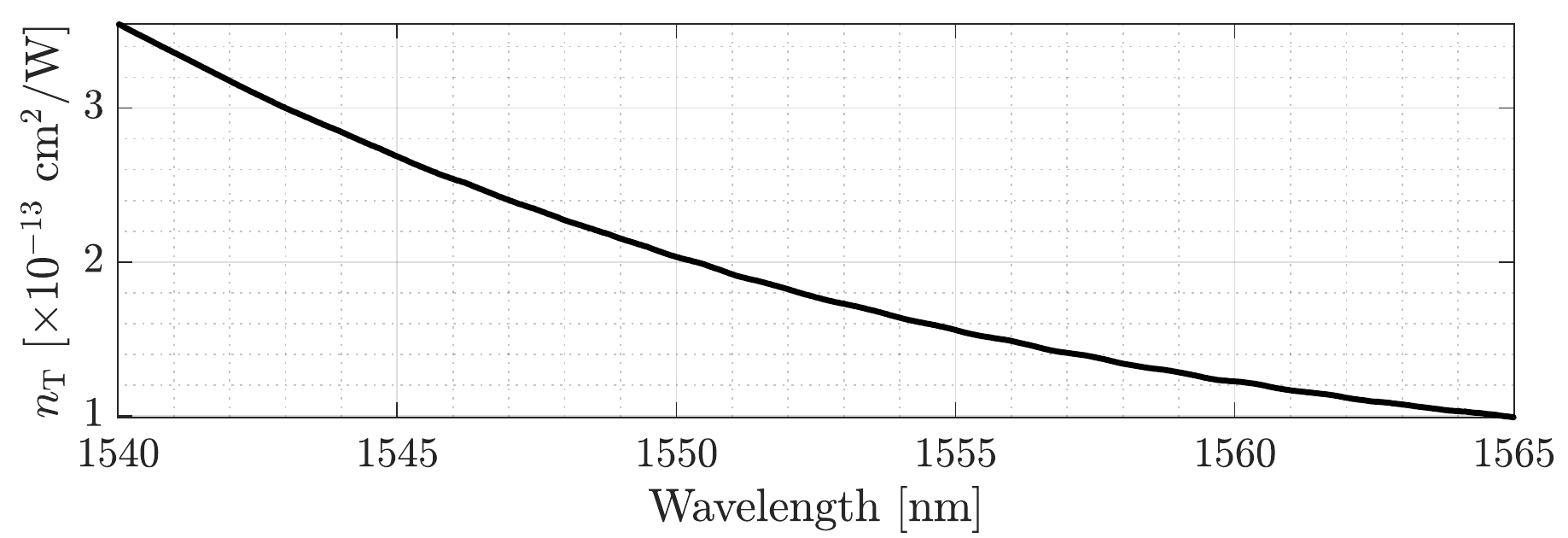}
	\caption{Thermo-optic nonlinear coefficient as a function of the wavelength.}
	\label{fig:nT}
\end{figure}

In this model we consider the following relationship between $n_{\rm T}$ and the propagation losses: $n_{\rm T}\propto 1-e^{-\alpha p}$, where $\alpha$ and $p$ are the propagation losses and the microresonator perimeter, respectively.
By comparing experimental and simulated spectra, we obtained an estimation of $n_{\rm T}$ reported in \figname\ref{fig:nT}.

\section*{Acknowledgement}

We acknowledge financial support from the Provincia Autonoma di Trento and the Q@TN initiative. S.B. acknowledges funding from the MIUR under the project PRIN PELM (20177 PSCKT). We thank F. Ramiro Manzano, A. Calabrese and H. M. Price for continuous and insightful exchanges and E. Moser for technical support.

\bibliographystyle{unsrt}
\bibliography{main.bib}

\end{document}